\documentclass[aps,prd,twocolumn,showpacs,preprintnumbers,nofootinbib]{revtex4}

\usepackage{graphicx}
\usepackage[FIGTOPCAP]{subfigure}
\usepackage{dcolumn}
\usepackage{bm}
\usepackage[normalem]{ulem}

\usepackage{amsmath}
\usepackage{epstopdf}
\usepackage{amsmath,amssymb}

\newcommand{\beq}{\begin{eqnarray}}
\newcommand{\eeq}{\end{eqnarray}}

\newcommand{\real}{{\sf I}\kern-.12em{\sf R}}
\newcommand{\comp}{{\sf I}\kern-.50em{\sf C}}
\newcommand{\unity}{{\sf I}\kern-.54em{\sf 1}}

\begin{document}

\title{
Phase structure of compactified $SU(N)$ gauge theories 
in magnetic backgrounds
}

\author{Massimo D'Elia}
\email{massimo.delia@unipi.it}
\affiliation{INFN - Sezione di Pisa, Largo Pontecorvo 3, I-56127 Pisa, Italy}
\affiliation{Dipartimento di Fisica dell'Universit\`a di Pisa, Largo Pontecorvo 3, I-56127 Pisa, Italy}

\author{Marco Mariti}
\email{marco.mariti@df.unipi.it}
\affiliation{INFN - Sezione di Pisa, Largo Pontecorvo 3, I-56127 Pisa, Italy}
\affiliation{Dipartimento di Fisica dell'Universit\`a di Pisa, Largo Pontecorvo 3, I-56127 Pisa, Italy}

\date{\today}

\begin{abstract}
We discuss the properties of non-abelian gauge theories formulated 
on manifolds with compactified dimensions and in the presence of 
fermionic fields coupled to magnetic backgrounds. 
We show that different phases may emerge, corresponding to different
realizations of center symmetry and translational invariance, depending 
on the compactification radius and on the magnitude of the magnetic
field. Our discussion focuses on the case of an
$SU(3)$ gauge theory in 4 dimensions with 
fermions fields in the fundamental representation, for which 
we provide some exploratory numerical lattice results. 
\end{abstract}


\maketitle

\section{Introduction}\label{sec:intro}

The purpose of this study is to investigate a class of phenomena
taking place in $SU(N)$ gauge theories with dynamical fermion
fields, when one of the space-time dimensions is compactified in the 
presence of an electromagnetic background coupled to the 
fermions. Such phenomena result from the coupling of the
gauge field holonomy to the background, through fermion
loops, leading to an entanglement between center and traslational
symmetries, which manifests itself through the presence of different
phases and phase transitions.

Center symmetry is known to play a fundamental role in determining
the phase diagram of pure gauge theories~\cite{thooft78}.
In that case, the action is symmetric under gauge transformations which are
periodic in the compactified direction, apart from a constant
element belonging to the center of the gauge group. 
In the lattice formulation that can be rephrased in terms of 
multiplication of all gauge links pointing in the compactified direction
and taken at a given slice orthogonal to it,
by  an element of the center of the gauge group 
$Z_{N}\equiv\left\{e^{i 2 k \pi/N}, k = 0, \dots  N - 1 \right\}$.
This symmetry can be exact or spontaneously broken;
the trace of the holonomy along the compactified 
direction, $L \equiv \rm Tr \exp (i \oint d x_\mu g A_\mu)$ 
(Wilson line or Polyakov loop), 
which gets multiplied by the corresponding 
center element, is a possible order parameter.
Its expectation 
value $\langle L \rangle$ becomes non-zero 
and proportional to a center element for 
small enough compactification radii,
due to the appearance of $N$ degenerate
vacua in the holonomy effective potential.
For a thermal compactification, the corresponding 
phase transition describes deconfinement~\cite{svet82}.

The presence of matter fields changes the picture substantially.
The covariant derivative in the 
fermion action introduces a direct coupling to the holonomy
around the compactified direction, which breaks center symmetry explicitly.
In particular, for fermions in the fundamental representation
and thermal boundary conditions (b.c.),
this coupling tends to favor
a real Wilson line,
so that the spontaneous breaking disappears\footnote{For periodic
b.c., positive real values of Wilson line are instead disfavored, and
a spontaneous breaking of the residual center symmetry is 
still possible~\cite{dh07,lpp07,ds09}.  
}.

\begin{figure}[t!]
\begin{center}
\includegraphics*[width=0.45\textwidth]{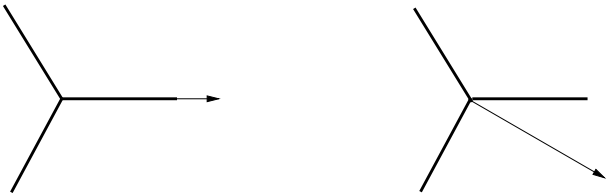}
\end{center}
\caption{The introduction of dynamical fermions breaks center symmetry
like an external field in a spin system, e.g., a
3-state Potts model for $SU(3)$. For standard thermal b.c.
the external field points along the real axis (left), while the 
introduction of an imaginary chemical potential rotates it (right).}
\label{fig0}
\end{figure}

An even more interesting phenomenology takes place
when an additional $U(1)$ background is coupled to 
matter fields, i.e. when the fermionic covariant derivative
is 
\beq
D_\nu = \partial_\nu + i \, g A^a_\nu T^a + i \, q a_\nu
\eeq
where $T^a$ are the $SU(N)$ generators
and $q$ is the coupling to the external $U(1)$ field $a_\mu$.
A well known example is that of an external imaginary chemical potential,
$\mu = i \mu_I$ in QCD at finite temperature.
In this case $q a_\nu = \mu_I \delta_{\nu\, 0}$, where 
$0$ is the Euclidean temporal direction, and the full holonomy entering
the fermion determinant is 
\beq
{\rm Tr} \exp \left(\oint d x_\mu i (g A_\mu + q a_\mu) \right) = 
L e^{i \mu_I/ T} \, .
\eeq
It is therefore $L \exp(i \mu_I / T)$ which tends to be oriented
along the real direction, i.e. in this case the effects of fermion fields
tends to align $L$ along $\exp(- i \mu_I / T)$, like an external field
whose direction in the complex plane is fixed by $\mu_I/T$. 
This is exemplified in Fig.~\ref{fig0}
for the case of $SU(3)$.
In the high $T$ phase, where the pure gauge contribution to the 
effective potential of the holonomy would tend to align it along a 
center element, that results in first order phase transitions
as $\mu_I/T$ crosses $\pi / N$ or odd multiples of it,
which are known as Roberge-Weiss transitions~\cite{rw}.

We are going to explore what happens when 
the $U(1)$ background is non-uniform.
To fix ideas, we will consider the case in which
a spatial dimension gets compactified in the presence of 
a background magnetic field,
notice however that the case in which the compactified dimension
has thermal b.c. is completely equivalent, since
the antiperiodic b.c. for fermions imply just a global
shift for the background field. Moreover, we will consider 
for simplicity the case in which all fermions have the same 
electric charge.

\begin{figure}[t!]
\includegraphics*[width=0.39\textwidth]{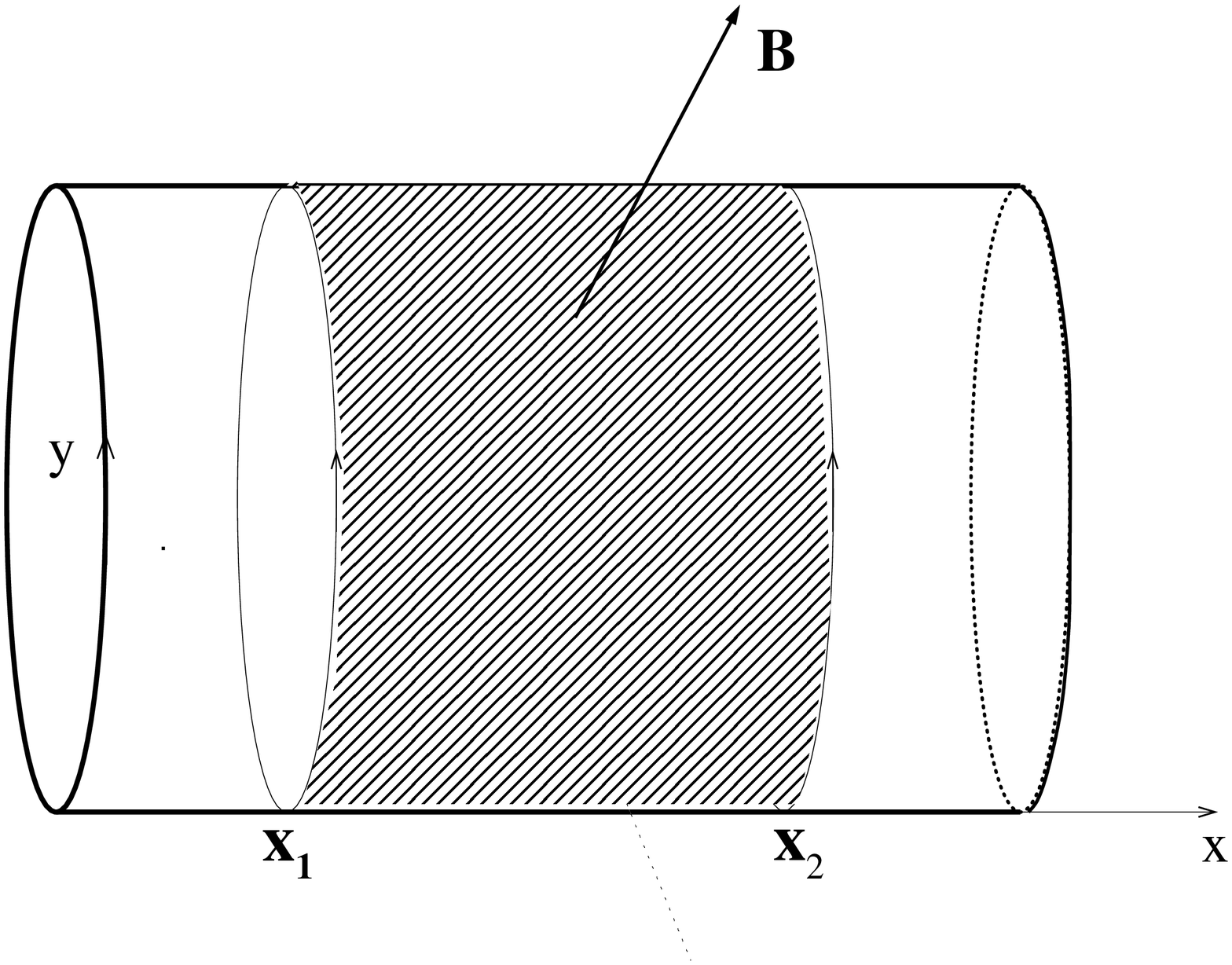}
\caption{Space dimension $y$ is compactified in the presence of a 
background field. The non-abelian holonomies sitting at $x_1$
and $x_2$ couple differently to dynamical fermions, depending
on the flux of the field across the shaded surface.}
\label{fig:torus}
\end{figure}

Let us consider the situation depicted in Fig.~\ref{fig:torus}:
direction $\mathbf{y}$ is compactified, with a compactification
length $L_c$,
in the presence of a
magnetic field orthogonal to the $\mathbf{x - y}$ plane.
The Wilson line sitting at $\mathbf{x_1}$ will couple to dynamical 
fermions of charge $q$ through a local phase factor,
i.e.~in the combination
$L(\mathbf{x}) e^{i q \oint d y a_y(\mathbf{x},y)} =
L(\mathbf{x}) e^{i \phi (\mathbf{x})}$:
such a coupling will tend to align the Wilson line along the 
center element closest to $e^{- i \phi(\mathbf{x})}$. 
However, since the phase factor depends on $\mathbf{x}$, 
it will tend to align
Wilson lines sitting at different values of the non-compactified 
coordinates along different center elements, i.e.~the $U(1)$ background 
field will induce, for small enough $L_c$, 
a structure of different center domains.

Whereas the value of a single phase factor is not physically relevant and gauge dependent, 
the phase difference between different points is. Indeed we have
\beq
e^{i (\phi(\mathbf{x_2}) - \phi(\mathbf{x_1}))}
&=& 
\exp \left(i q \oint d y (a_y(\mathbf{x_2},y) - a_y(\mathbf{x_2},y))
\right) 
\nonumber \\
&=&
e^{i q \Phi_{\mathbf B}}
\label{fluxeq} 
\eeq
where $\Phi_{\mathbf B}$ is the total magnetic 
field flux going through 
the shadowed surface in the figure.
Despite the simplified situation in Fig.~\ref{fig:torus}, it is
easy to realize that
the value of this flux is, for any magnetic field distribution, 
independent of the particular shape of the surface in the 
non-compactified directions, i.e. it is a property of the
points $\mathbf{x_1}$ and $\mathbf{x_2}$ only.
Therefore, modulo a global center rotation, the structure of 
center domains that tends to be formed is a unique property of 
the magnetic field distribution.

However,
the fact that such a structure actually forms is non-trivial, since
the formation of center domains 
implies the presence of interfaces separating them, which has a cost
in terms of energy. The actual structure will depend on the 
balance between the energy spent in creating center 
interfaces and the energy spent in keeping the holonomy 
in a locally wrong vacuum: the former is a function of the interface tension 
and of the density of interfaces, which depends on the magnetic
field strength, the latter is a function of the holonomy 
effective potential. Since both 
the 
interface tension and the effective potential 
are functions of $L_c$, one may expect 
that different phases, corresponding to 
different center domain structures, are crossed as the 
compactification radius shrinks, with a corresponding
presence of
phase transitions and metastable
states.

To discuss that more in detail,
let us assume that $y$ in Fig.~\ref{fig:torus}
is the spatial direction of
a 4-dimensional (4D) gauge theory and  that, for simplicity,
the background field $F_{xy} = B$ is uniform and constant. 
We will compare
two extreme situations: that in which all center domains
are actually formed, i.e. the holonomy is in the correct ``local vacuum''
everywhere, and that in which the holonomy stays in the same center 
sector everywhere, without forming any interface. 
In the first case, making reference 
to Fig.~\ref{fig:torus}, the number of interfaces, $N_{int}$, is given by
the different center sectors spanned by the local phase between
$x_1$ and $x_2$, i.e. 
\beq
N_{int} = q \Phi_B/(2 \pi /N) = q B L L_c N /2 \pi\, 
\label{defB0}
\eeq
where $L = |x_2 - x_1|$,
while in the second case one must keep the holonomy in the wrong 
center sector for a fraction $(N  - 1)/N$ of the region
between $x_1$ and $x_2$.

In the limit of asymptotically small $L_c$, we can recover
perturbative results obtained in thermal field theory, where 
the role of the compactified direction is played by the Euclidean time 
direction and $T = 1/L_c$. The interface tension 
(i.e. the energy per unit interface area)
is proportional to $L_c^{-3}\, \log (1/L_c)$~\cite{tension}, 
and the energy density 
spent to keep the holonomy in the wrong vacuum is proportional 
to $L_c^{-4}$~\cite{rw}. 
Without considering a common factor related to the 
integration over the non-compactified directions orthogonal to $x$, 
the energy spent
to create all possible interfaces between $x_1$ and $x_2$ is then 
proportional to $q B L L_c^{-2} \log (1/L_c)$, while the energy spent
to maintain the holonomy in the same center sector, without creating
any interface, is proportional to
$L L_c^{-4}$. 
It is clear that the first situation is surely favored, at fixed
magnetic field, for small enough $L_c$ and, at fixed $L_c$, 
for small enough $B$. For intermediate values of $L_c$
and/or $B$, the lowest energy configuration might correspond
to a partial formation of the center domain structure,
so that various phase transitions can be crossed as 
the two quantities change.
Given the power law dependence on $L_c$, a similar behavior
is expected also when $L_c$ is changed at fixed total flux,
i.e. if $B$ is scaled proportionally to $1/L_c$.

\begin{figure}[t!]
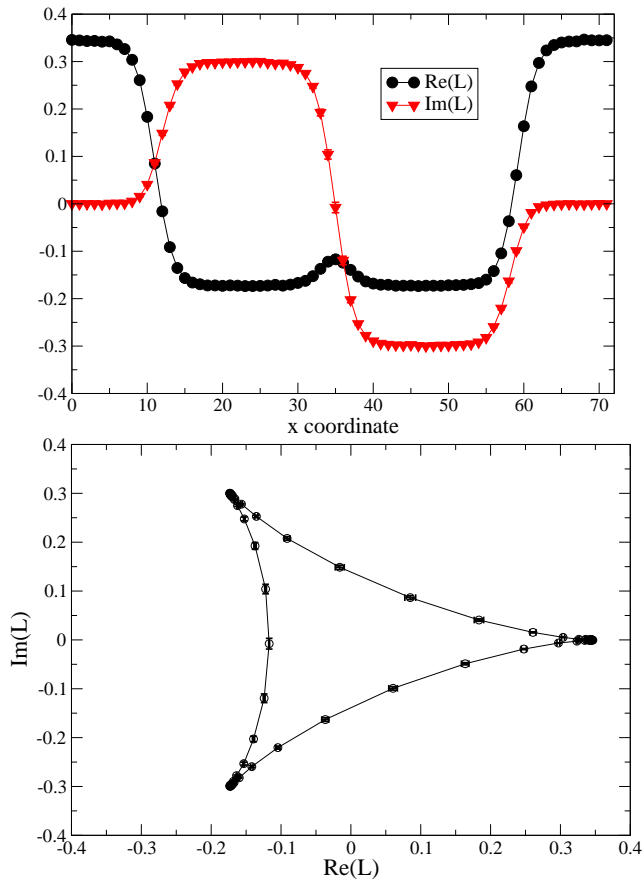

\begin{center}
\includegraphics*[width=0.43\textwidth]{spatialdep_lc4_lx72_b1.eps}
\includegraphics*[width=0.47\textwidth]{complex_lc4_lx72_b1.eps}
\end{center}
\caption{
Local average of the real and imaginary part of 
the Wilson line as a function of $x$ (up) and in the complex plane (down),
for $L_x = 72$, $L_c = 4$ and $b = 1$ (see Eq.~(\ref{defB})).
All predicted center domains are explored as $x$ changes,
with the associated interfaces separating them.}
\label{fig1}
\end{figure}

\begin{figure}[t!]
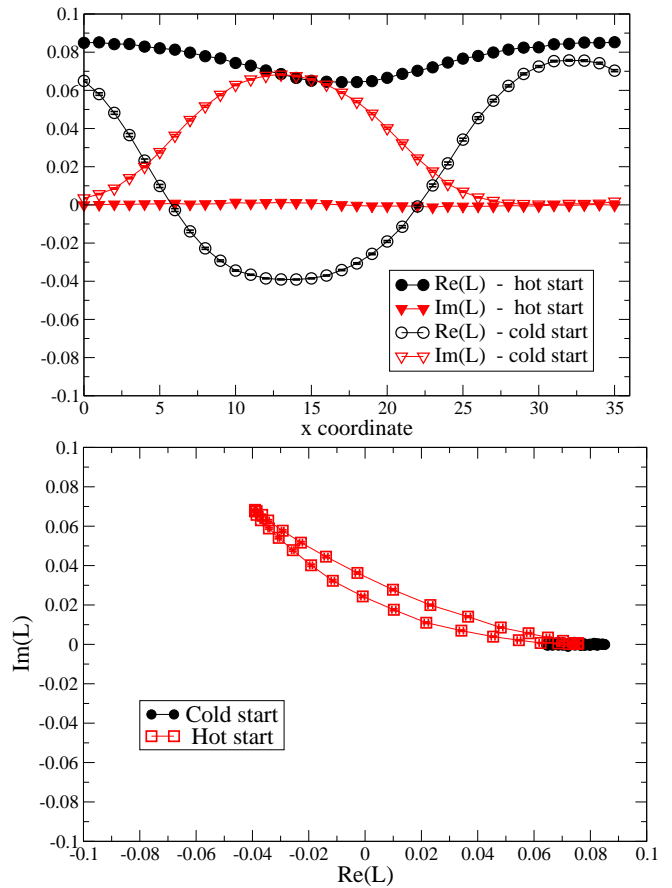

\begin{center}
\includegraphics*[width=0.44\textwidth]{spatialdep_lc8_lx36_b1.eps}
\includegraphics*[width=0.48\textwidth]{complex_lc8_lx36_b1.eps}
\end{center}
\caption{As in Fig.~\ref{fig1}, for $L_x = 36$, $L_c = 8$, $b = 1$
(corresponding to the same $B$ of Fig.~\ref{fig1}),
and two different simulations, starting from
random (hot) or unit (cold) gauge links.
In one case (cold start) a single center domain is formed, in the other a phase with 
two center domains emerges.
In both cases, the 
center-translational symmetry is spontaneously broken.
}
\label{fig2}
\end{figure}

We notice that, in the case of a constant and uniform 
magnetic background, an exact center-translational symmetry appears, since 
an elementary center trasformation
can be exactly reabsorbed by 
a traslation along $x$ by $2 \pi / (q B L_c N)$.
This discrete symmetry can be
either exactly realized or spontaneously broken.
In the first case, 
after each translation by $2 \pi / (q B L_c N)$ the holonomy rotates by 
$-2 \pi /N$, and the spatial average of the Wilson line is exactly zero.
In the second case, 
the holonomy fails to rotate, because interfaces 
cost too much and it is more convenient to stay in the 
false vacuum somewhere; as a consequence, the spatial average of the Wilson
line, which serves as a non-local order parameter, could be non-zero.

\section{Numerical simulations}

In order to test this scenario,
we have performed  
numerical simulations of a 4D $SU(3)$
gauge theory, with two degenerate and equally charged 
dynamical flavors in the fundamental representation,
adopting the Rational Hybrid Monte Carlo algorithm~\cite{rhmc}
and the code developed in Ref.~\cite{gpupaper}.
The theory has been discretized on a periodic 4D torus, 
with a constant and uniform 
magnetic field orthogonal to the $x-y$ plane and the $y$ direction
significantly shorter than the others, as in Fig.~\ref{fig:torus}.
We have considered a standard rooted staggered discretization of the theory. The 
partition function reads:
\beq
Z \equiv \int \mathcal{D}U e^{-S_{G}} 
\det D^{1\over 2} [U,q]
\:
\label{partfun1}
\eeq
\begin{eqnarray}
D^{(q)}_{i,j} \equiv a m \delta_{i,j} 
&+& {1 \over 2} \sum_{\nu=1}^{4} \eta_\nu(i) \left(
u_\nu^{(q)}(i)\ U_{\nu}(i) \delta_{i,j-\hat\nu}
\right. \nonumber \\ &-& \left.
u^{*(q)}_\nu{(i - \hat\nu)}\ U^{\dag}_\nu{(i-\hat\nu)} \delta_{i,j+\hat\nu} 
\right) \:
\label{fmatrix1}
\end{eqnarray}
$\mathcal{D}U$ is the integration over $SU(3)$ gauge link
variables, $S_G$ is the standard Wilson plaquette pure gauge action,  
$i$, $j$ are lattice site indexes,
$\eta_{\nu}(i)$ are the staggered
phases. The $U(1)$ phases $u_\mu^{(q)}(i)$ appearing in the fermion
matrix are chosen so as to reproduce a uniform magnetic field across
the $x-y$ plane, which, as a consequence of the periodic b.c.,
is quantized
according to~\cite{bound1,bound2,bound3,wiese} 
\beq
q B = {2\pi b}/{(L_x L_y a^2)}
\label{defB}
\eeq
where $b$ is an integer.
It is easy to check that the number of different center
sectors which should be crossed when moving along the $x$ 
direction is exactly equal to $b N$.

We have worked in a fixed cut-off scheme, 
setting the inverse gauge coupling $\beta = 6/g_0^2$ and 
the bare quark mass respectively to 
$\beta = 6.2$ and $am = 0.01$ in all simulations performed.
We have considered $L_x \times L_y \times L_z \times L_t$ lattices,
fixing $L_x = L_t = 24$,
then tuning $L_y = L_c$ to change the compactification radius, and 
$L_x$ and $b$ to change the magnetic background at fixed $L_c$.
Such bare values correspond roughly
to a pion mass of the order of the  $\rho$ mass~\cite{blum}. 
For all explored values of the compactification radius, the corresponding 
thermal system at zero background field is in the deconfined phase.

\begin{figure}[t!]
\begin{center}
\includegraphics*[width=0.45\textwidth]{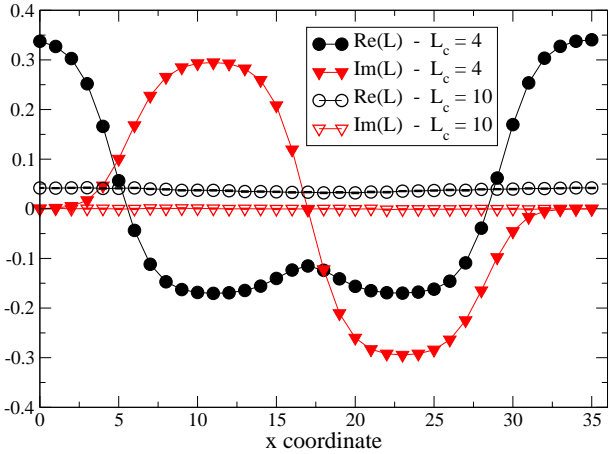}
\end{center}
\caption{Local average of the real and imaginary part of 
the Wilson line for $L_x = 36$ and two different compactifications,
$L_c = 4, 10$, keeping the total magnetic flux unchanged
($b = 1$). 
}
\label{fig3}
\end{figure}

In Fig.~\ref{fig1} we show results obtained for the real and imaginary part
of the Wilson line for simulations with 
$L_y = 4$, $L_x = 72$ and $b = 1$. Expectation values are reported
both as a function of $x$ and in the complex plane: the formation 
of the predicted three center domains, separated by three
interfaces, is clearly visible, and the center-translational symmetry
is realized exactly, i.e. the system is globally center-symmetric.
However, as we increase $L_c$ while keeping the magnetic field,
hence $b/(L_c L_x)$, fixed, the situation changes.

In Fig.~\ref{fig2} we report results obtained for $L_c = 8$.
In this case two different phases are found, depending on the 
starting configuration of the Monte-Carlo simulation. In both
of them the global center symmetry is spontaneously broken:
in one phase the system chooses a single center domain, with no
interface, similarly to a standard thermal
system in the high-$T$ regime, so
we can name it ``deconfined phase''; in the other instead two center domains
are formed, with the corresponding separating interfaces and, due to 
the characteristic
shape in the complex plane (see Fig.~\ref{fig2}), we name it 
``banana phase''.
For $L_c = 6$ one finds that the global center symmetry is exact, while
for $L_c > 8$ only the deconfined phase survives. The 
metastability found for $L_c = 8$ is a clear suggestion that the different
phases are separated by strong first order transitions.

\begin{figure}[t!]
\begin{center}
\includegraphics*[width=0.48\textwidth]{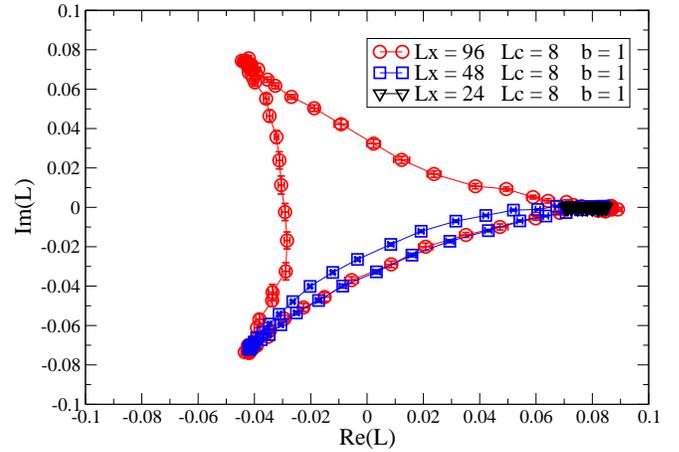}
\end{center}
\caption{Local average of the real and imaginary part of 
the Wilson line in the complex plane, for $L_c = 8$ and three different
values of the magnetic field. A different phase is explored in each case.
}
\label{fig4}
\end{figure}

\begin{figure}[t!]
\begin{center}
\includegraphics*[width=0.48\textwidth]{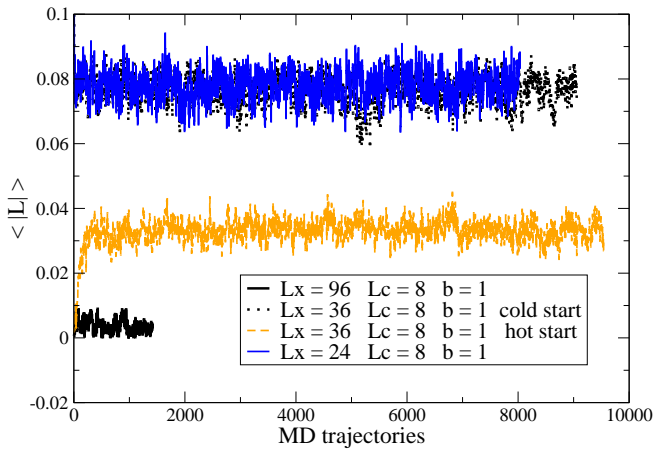}
\end{center}
\caption{Monte-Carlo time histories of the 
average modulus of the spatially averaged Wilson line,
which is an order parameter for the realization of the 
center-translational symmetry, for simulations at fixed $L_c$
and variable magnetic field.} 
\label{fig5}
\end{figure}

A similar pattern takes place if one changes $L_c$ while keeping a fixed
magnetic flux, i.e. by scaling $B \propto 1/L_c$.
This is visible in Fig.~\ref{fig3}, which shows two 
different compactifications, $L_c = 4$ and 10, where the flux 
is the same as for the $L_c = 8$ case in Fig.~\ref{fig2}.

Finally, in Fig.~\ref{fig4} we show a set of results in which
the compactification radius is kept fixed and one changes the magnetic field.
As expected, as $B$ increases the 
system moves from the phase with an exact global center symmetry, to 
the banana phase and, finally, to the deconfined phase; in all 
showed examples the phases are stable, i.e. they are found independently
of the starting configuration.

A non-local order parameter for the realization of the center-translational
symmetry is the spatial average of the Wilson line over all non-compactified directions.
Its time history is reported in Fig.~\ref{fig5} for some of the cases
discussed above and, in particular, for the metastable
case reported in Fig.~\ref{fig2}.

\section{Conclusions}

To summarize, we have discussed how the compactification 
of a non-abelian gauge theory in the presence
of a $U(1)$ background field is accompanied by the 
formation of a  structure of center domains, 
dictated by the dynamics of the holonomy, for asymptotically
small values of $L_c$. As $L_c$ increases, the energetically favorable
structure can change, leading to different phases 
characterized by a reduced number of domains; such phases
are likely separated by first order transitions, leading 
to the formation of metastable states. 
A similar behavior is found as 
$B$ increases at fixed $L_c$.

We have focused on the simplified case of an $SU(3)$ gauge
group and of fermions with degenerate
charges. Of course, if the electric charges are different
and/or for different gauge groups, 
the structure of center domains, dictated by the local minima of 
the holonomy, can be different, because of the competing contributions
from fermions with different electric charges. However, the general 
picture, in particular the appearance of different phases and metastable
states as $L_c$ and/or $B$ change, will be qualitatively similar. 
Also the addition of more non-compactified space dimensions should 
not change the scenario, which theorefore
could be of interest for theories with extradimensions which
get compactified in the presence of background fields.

\acknowledgements

We thank Claudio Bonati, Michele Mesiti and Francesco Negro for 
very useful discussions.
Numerical simulations have been performed 
on a GPU farm located at the INFN Computer Center in Pisa and on the
QUONG cluster in Rome.


\begin{thebibliography}{99}


\bibitem{thooft78} 
  G.~'t Hooft,
  Nucl.\ Phys.\ B {\bf 138}, 1 (1978).

\bibitem{svet82} 
  B.~Svetitsky and L.~G.~Yaffe,
  Nucl.\ Phys.\ B {\bf 210}, 423 (1982).
  doi:10.1016/0550-3213(82)90172-9

\bibitem{dh07} 
  T.~DeGrand and R.~Hoffmann,
  JHEP {\bf 0702}, 022 (2007)
  [hep-lat/0612012].

\bibitem{lpp07}
B.~Lucini, A.~Patella and C.~Pica,
  Phys.\ Rev.\ D {\bf 75}, 121701 (2007)
  [hep-th/0702167].


\bibitem{ds09} 
  M.~D'Elia and F.~Sanfilippo,
  Phys.\ Rev.\ D {\bf 80}, 111501 (2009)
  [arXiv:0909.0254 [hep-lat]].

\bibitem{rw}
A.~Roberge, N.~Weiss, Nucl. Phys. B {\bf 275}, 734 (1986).


\bibitem{tension} 
  T.~Bhattacharya, A.~Gocksch, C.~Korthals Altes and R.~D.~Pisarski,
  Phys.\ Rev.\ Lett.\  {\bf 66}, 998 (1991); 
Nucl.\ Phys.\ B {\bf 383}, 497 (1992).

\bibitem{bound1}
G.~'t Hooft,
  Nucl.\ Phys.\ B {\bf 153}, 141 (1979).
%
%
\bibitem{bound2} 
  J.~Smit and J.~C.~Vink,
  Nucl.\ Phys.\ B {\bf 286}, 485 (1987).
%
%
\bibitem{bound3} 
  P.~H.~Damgaard and U.~M.~Heller,
  Nucl.\ Phys.\ B {\bf 309}, 625 (1988).
%
%
\bibitem{wiese}
  M.~H.~Al-Hashimi and U.~J.~Wiese,
  Annals Phys.\  \textbf{324}, 343 (2009).

\bibitem{rhmc}
  M.~A.~Clark and A.~D.~Kennedy,
  Nucl.\ Phys.\ Proc.\ Suppl.\  {\bf 129}, 850 (2004) [hep-lat/0309084];
Phys.\ Rev.\ Lett.\  {\bf 98}, 051601 (2007) [hep-lat/0608015].



\bibitem{gpupaper} 
  C.~Bonati, G.~Cossu, M.~D'Elia and P.~Incardona,
  Comput.\ Phys.\ Commun.\  {\bf 183}, 853 (2012)
  [arXiv:1106.5673 [hep-lat]].

\bibitem{blum}
T.~Blum, L.~Karkkainen, D.~Toussaint and S.~A.~Gottlieb,
  Phys.\ Rev.\ D {\bf 51}, 5153 (1995)
  [hep-lat/9410014].


\end{thebibliography}
\end{document}